\def \SAIT #1 #2 {{\em Mem.\ Soc.\ Astron.\ It.\/} {\bf #1}, #2}
\def \MESS #1 #2 {{\em The Messenger\/} {\bf #1}, #2}
\def \ASTRNACH #1 #2 {{\em Astron. Nach.\/} {\bf #1}, #2}
\def \AAP #1 #2 {{\em Astron. Astrophys.\/} {\bf #1}, #2}
\def \AAL #1 #2 {{\em Astron. Astrophys. Lett.\/} {\bf #1}, L#2}
\def \AAR #1 #2 {{\em Astron. Astrophys. Rev.\/} {\bf #1}, #2}
\def \AAS #1 #2 {{\em Astron. Astrophys. Suppl. Ser.\/} {\bf #1}, #2}
\def \AJ #1 #2 {{\em Astron. J.\/} {\bf #1}, #2}
\def \ANNREV #1 #2 {{\em Ann. Rev. Astron. Astrophys.\/} {\bf #1}, #2}
\def \APJ #1 #2 {{\em Astrophys. J.\/} {\bf #1}, #2}
\def \APJL #1 #2 {{\em Astrophys.. J. Lett.\/} {\bf #1}, L#2}
\def \APJS #1 #2 {{\em Astrophys. J. Suppl.\/} {\bf #1}, #2}
\def \APSS #1 #2 {{\em Astrophys. Space Sci.\/} {\bf #1}, #2}
\def \ASR #1 #2 {{\em Adv. Space Res.\/} {\bf #1}, #2}
\def \BAIC #1 #2 {{\em Bull. Astron. Inst. Czechosl.\/} {\bf #1}, #2}
\def \JSQRT #1 #2 {{\em J. Quant. Spectrosc. Radiat. Transfer\/} {\bf #1},#2}
\def \MN #1 #2 {{\em Mon. Not. R. Astr. Soc.\/} {\bf #1}, #2}
\def \MEM #1 #2 {{\em Mem. R. Astr. Soc.\/} {\bf #1}, #2}
\def \PLR #1 #2 {{\em Phys. Lett. Rev.\/} {\bf #1}, #2}
\def \PASJ #1 #2 {{\em Publ. Astron. Soc. Japan\/} {\bf #1}, #2}
\def \PASP #1 #2 {{\em Publ. Astr. Soc. Pacific\/} {\bf #1}, #2}
\def \NAT #1 #2 {{\em Nature\/} {\bf #1}, #2}

\newcommand{\Omegachi}{\Omega_{\chi}}
\newcommand{\tb}{\tan\beta}

\newcommand{\sign}{\rm{sgn}}
\newcommand{\mgaugino}{M_{1/2}}

\newcommand{\be}{\begin{equation}}
\newcommand{\ee}{\end{equation}}
\newcommand{\etal}{{\em et al.}}

\def\picture #1 by #2 (#3){
  \vbox to #2{
    \hrule width #1 height 0pt depth 0pt
    \vfill
    \special{picture #3} 
    }
  }
\def\scaledpicture #1 by #2 (#3 scaled #4){{
  \dimen0=#1 \dimen1=#2
  \divide\dimen0 by 1000 \multiply\dimen0 by #4
  \divide\dimen1 by 1000 \multiply\dimen1 by #4
  \picture \dimen0 by \dimen1 (#3 scaled #4)}
  }
\def\boxit#1{$$\vbox{\hrule\hbox{\vrule\kern3pt
     \vbox{\kern3pt \hbox to 13cm{\hfil }\vskip
     #1cm\kern3pt} \kern3pt \vrule}
     \hrule}$$}

\documentclass{article}
\usepackage{FPSpro,epsfig,graphicx}
\begin{document}
\input epsf.sty
\title{ GAMMA RAY ASTROPARTICLE PHYSICS WITH GLAST}
\author{Aldo Morselli  \\
{\em Dept. of Physics,  Univ. of Roma ''Tor Vergata''  and INFN Roma 2, Italy}}
\maketitle
\baselineskip=11.6pt
\begin{abstract}
The  energy domain between 10 MeV and hundreds of GeV is an essential one for the multifrequency study of 
extreme astrophysical sources. The understanding of spectra of
detected gamma rays is necessary for developing models for acceleration, emission, absorption and propagation
of very high energy particles at their sources and in space. After the end of EGRET on board the Compton Gamma Ray Observatory
this energy region is not covered by any other experiment, at least up to 50 GeV where ground Cerenkov telescopes are beginning 
to take data. 
Here we will review the status of the space experiment GLAST that will fill this energy region from March 2006   with particular
emphasis at the connection with all the other ground and space planned experiments and at the contribution of GLAST
 to particle physics.
\end{abstract}

\baselineskip=14pt

\section{GLAST}

\begin{figure}[ht]
\vspace{0cm}
    \epsfig{file=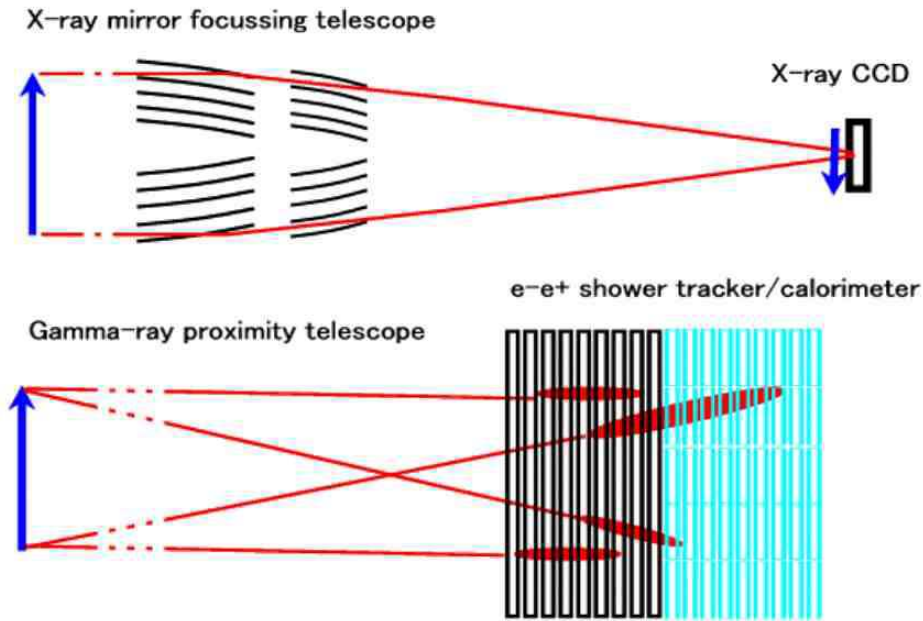,width=1.1\textwidth,angle=0,clip=}
\caption{\label{diffxg1} \it Detector Technology: X-ray versus Gamma-ray. }
\end{figure}

\begin{figure}
    \epsfig{file=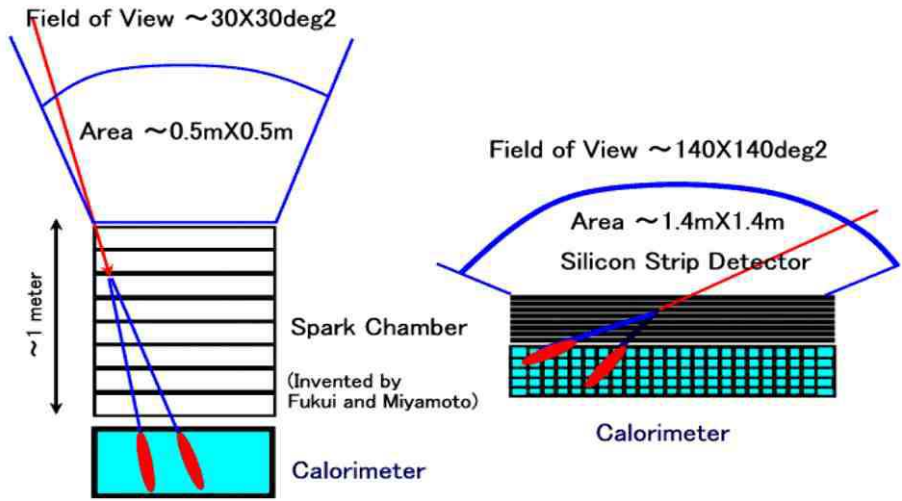,width=1\textwidth,angle=0,clip=}
\caption{\label{diffglastegr1} \it EGRET(Spark Chamber) versus GLAST (Silicon Strip
Detector).}
\vspace{-0.2cm}
\end{figure}

The techniques for the detection of gamma-rays in the pair production regime energy range are very different from the X-ray detection
ones. For X-rays detection focusing is possible and this permits large effective area,
excellent energy resolution, very low background. For gamma-rays  no focusing is  possible and this means
limited effective area,  moderate energy resolution, high background but a wide field of view (see figure~\ref{diffxg1}).
This possibility to have a wide field of view is enhanced now, in respect to EGRET, with the use of silicon detectors, that allow 
a further increase of the ratio between height and width (see fig.\ref{diffglastegr1}), essentially for two reasons: a) an increase of
the  position resolution that allow a decrease of the distance between the planes of the tracker without affect the angular resolution,  b)
the  possibility to use the silicon detectors themselves for the trigger of an events, with the elimination of the Time of Flight system, that
require some height. 


\begin{figure}
\vspace{0cm}
    \epsfig{file=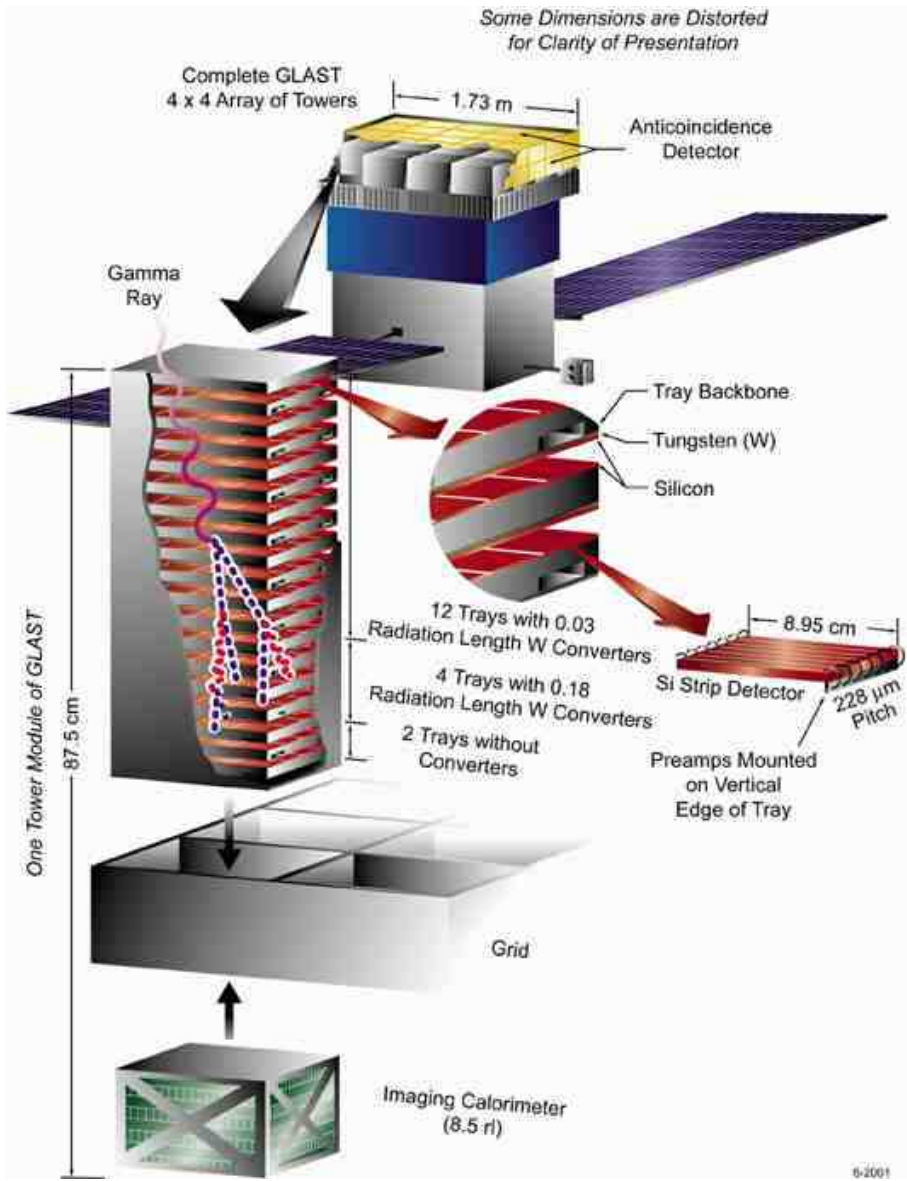,width=1\textwidth,angle=0,clip=}
\caption{\label{glastscheme} \it The GLAST instrument, exploded to show
the detector layers in a tower, the stacking of the CsI
logs in the calorimeter, and the integration of the subsystems.}
\end{figure}

The Gamma-ray Large Area Space Telescope (GLAST)\cite{glast},  has
been selected by NASA  as a  mission involving an international collaboration of  particle physics and astrophysics 
communities from
the United States, Italy, Japan, France and Germany  for a launch in the first half of 2006. 
The main scientific objects are
the study of all gamma ray sources such as blazars, gamma-ray bursts,  supernova
remnants, pulsars, diffuse radiation, and unidentified high-energy sources.
Many years of refinement has led to the configuration of the apparatus shown in figure~\ref{glastscheme},
where one can see the  4x4 array of identical towers each formed by:
$\bullet $   Si-strip Tracker Detectors and converters arranged in 18 XY tracking planes for the measurement
of the photon direction.
$\bullet $ Segmented array of CsI(Tl) crystals for the measurement the photon energy.
$\bullet $ Segmented Anticoincidence  Detector (ACD).
The main characteristics
are an  energy range between	20 MeV and 300 GeV, 
a field of view of $\sim$	3  sr,  an energy resolution	of $\sim$ 5\% at 1 GeV, 
a point source sensitivity of  2x10$^{-9}$ (ph~cm$^{-2}$~s$^{-1}$)     at  0.1 GeV,          
an event deadtime	 of 20 $\mu s$ and a peak effective area  of  10000 cm$^2$,  
for a required power	of   600 W and a payload weight	of  3000 Kg.

 The list of the
people and  the Institution involved in the collaboration together with the on-line
status of the project is available at {\sl http://www-glast.stanford.edu}. 
A  description of the apparatus  can be found in\cite{bellazzini}.
\begin{figure}
\vspace{0cm}
    \epsfig{file=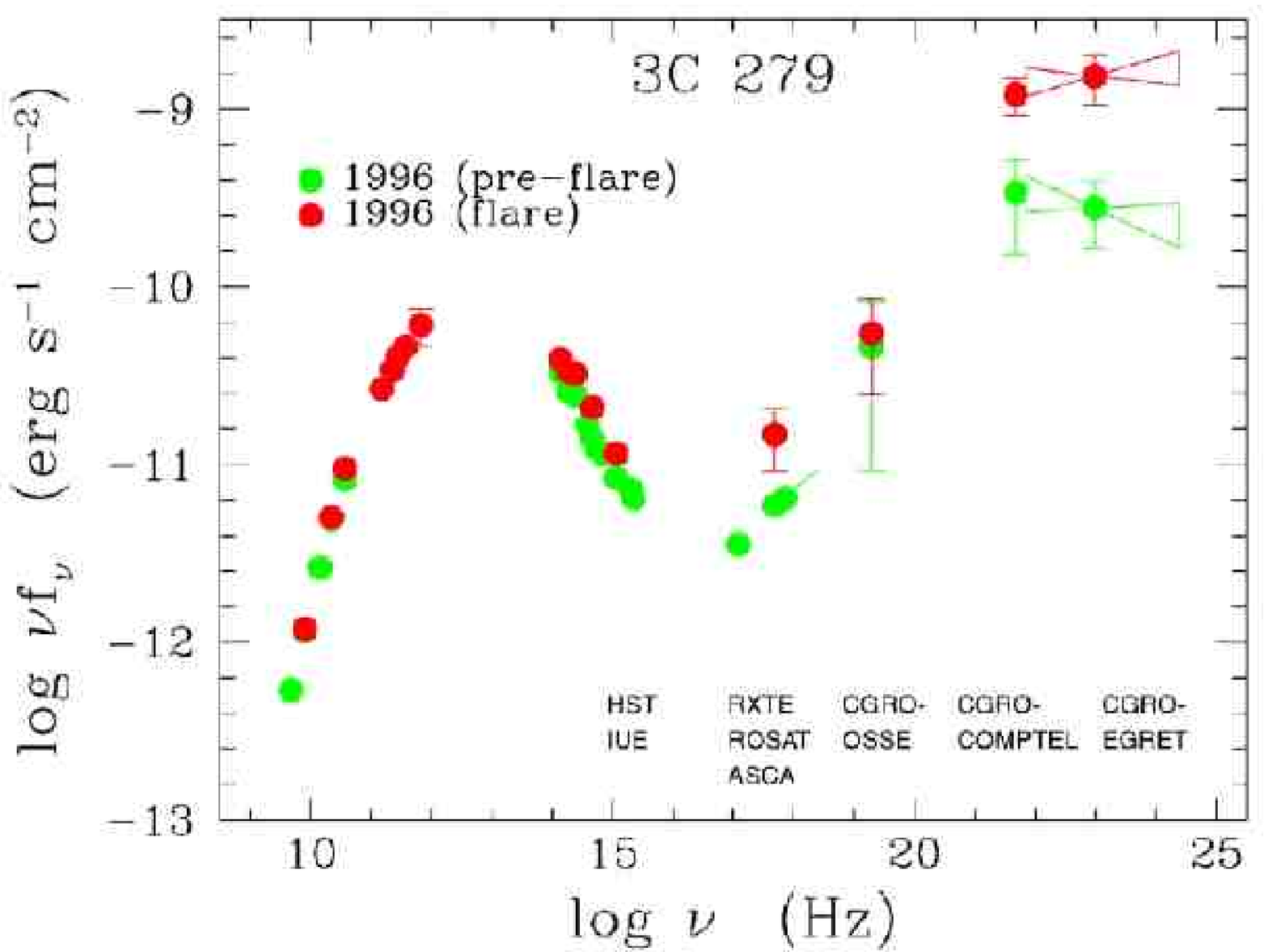,width=0.8\textwidth,angle=0,clip=}
\caption{\label{3c279a} \it Spectral energy distributions of the quasars 3C 279 during flaring state (in red) and non flaring state (in
green).}
\end{figure}
\begin{figure}
\vspace{0cm}
    \epsfig{file=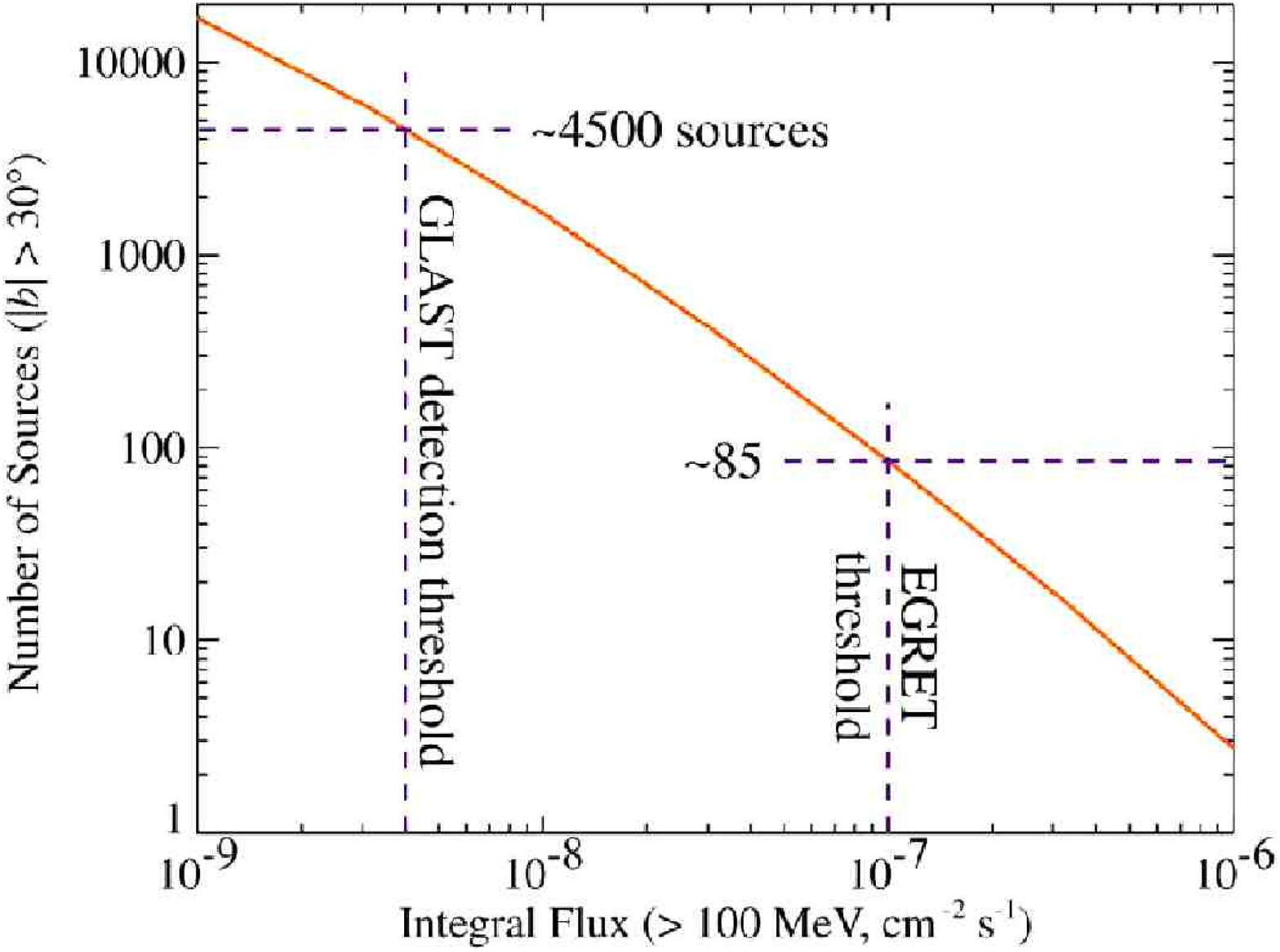,width=0.8\textwidth,angle=0,clip=}
\caption{\label{lognlogs1} \it  Estimate of the number of AGNs that GLAST will detect at high latitude in a 2 
year sky survey compared to EGRET's approximate detection limit ~$^{4)}$.}
\vspace{0.5cm}
    \epsfig{file=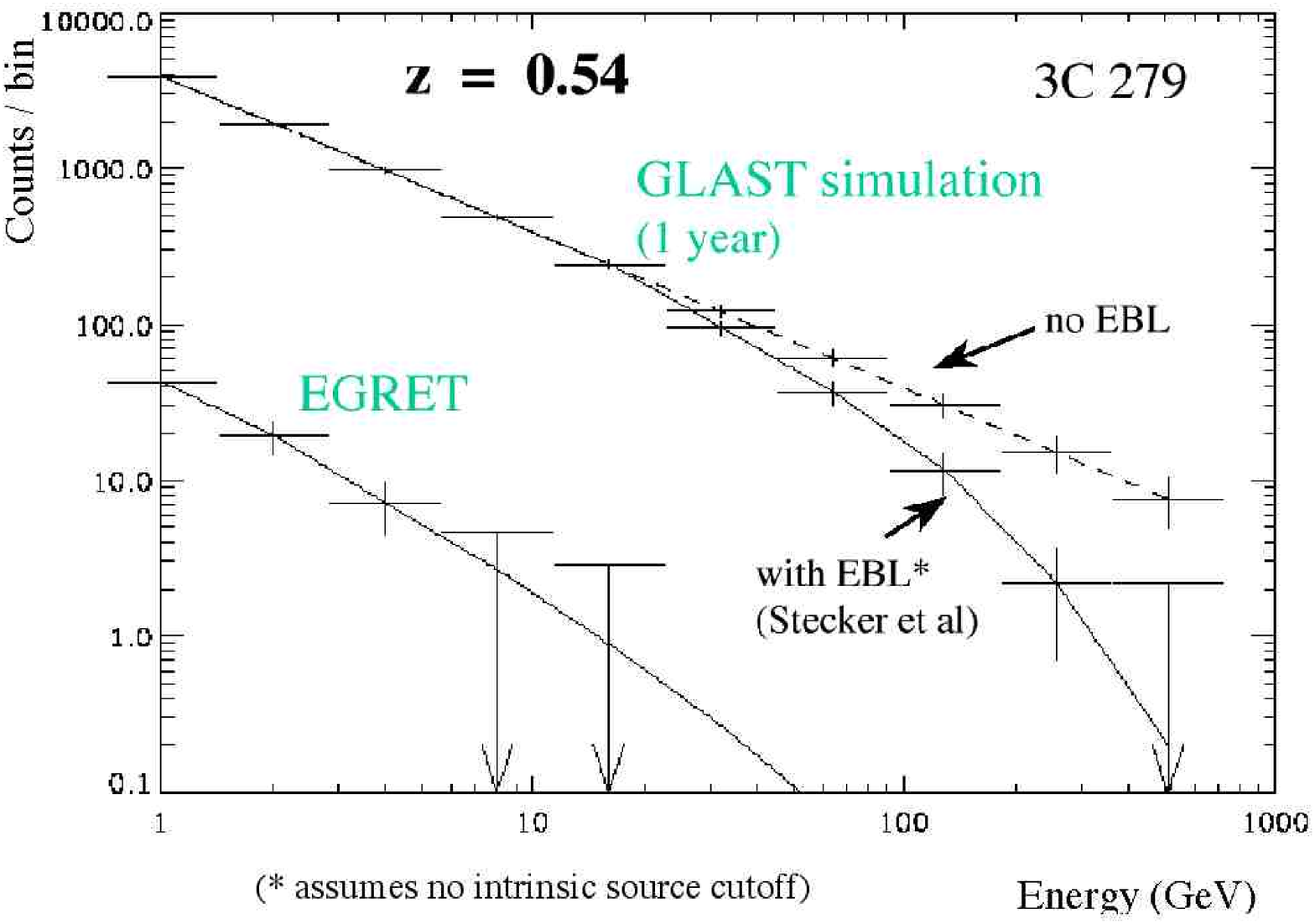,width=0.8\textwidth,angle=0,clip=}
\caption{\label{attglast2} \it Number of photons detected by EGRET from 3C279 and the number expected 
with GLAST in the case of extragalactic background light attenuation and without attenuation ~$^{5)}$.}
\end{figure}
\subsection{ Active Galactic Nuclei }

Before EGRET, 3C 273 was the only active galactic
nucleus (AGN) known to emit high-energy gamma
rays. Now we known that there is an entire class of active galaxies that
probably represent the largest class of high energy gamma-ray emitters:
the blazars. Blazars are flat radio spectrum, active galactic nuclei, or
AGN, whose members include BL Lac objects and highly polarized and
optically violently variable quasars that often emits more in gamma-ray
than in any other frequencies (see figure~\ref{3c279a}). For a review on AGNs see reference\cite{celotti}.
GLAST will dramatically extend the number of observed AGNs, as well as the energy range over which they can be observed. Indeed, GLAST might be called
the "Hubble Telescope" of gamma-ray astronomy as it will be able to observe AGN sources to z~$\sim$~4  and beyond, if such objects actually existed at
such early times in the universe.  Figure~\ref{lognlogs1} shows the so called Log N versus Log S distribution, where N is the number of
sources and S is source flux for $E_\gamma > 100$ MeV, for AGN. The curve is extrapolated from EGRET data and an AGN model of the diffuse
gamma-ray background based on the assumption that AGN sources follow a luminosity function similar to flat spectrum radio quasars.
Extrapolation from EGRET AGN detections projects that about 5,000 AGN sources will be detected in a 2 year cumulative scanning mode
observation by GLAST, as compared to the 85  that have been observed by EGRET in a similar time interval.  This large number of AGN's
covering a redshift range from z $\sim$ 0.03  up to  z $\sim$ 4 will allow to disentangle   an intrinsic
cutoff effect, i.e., intrinsic to the source, from a cut-off derived from the interaction with the extra
galactic background light, or EBL. Only by observing many examples of AGN, and over a wide range of
redshifts, one can  hope to untangle these two possible sources of cutoff.  In figure~\ref{attglast2} is
shown the number of photons detected by EGRET from 3C279 and the number expected  with GLAST in the case of
extragalactic background light attenuation and without attenuation. Determination of the EBL can provide
unique information on the formation of galaxies at early epochs, and will test models for structure formation
in the Universe.
\begin{figure}
    \epsfig{file=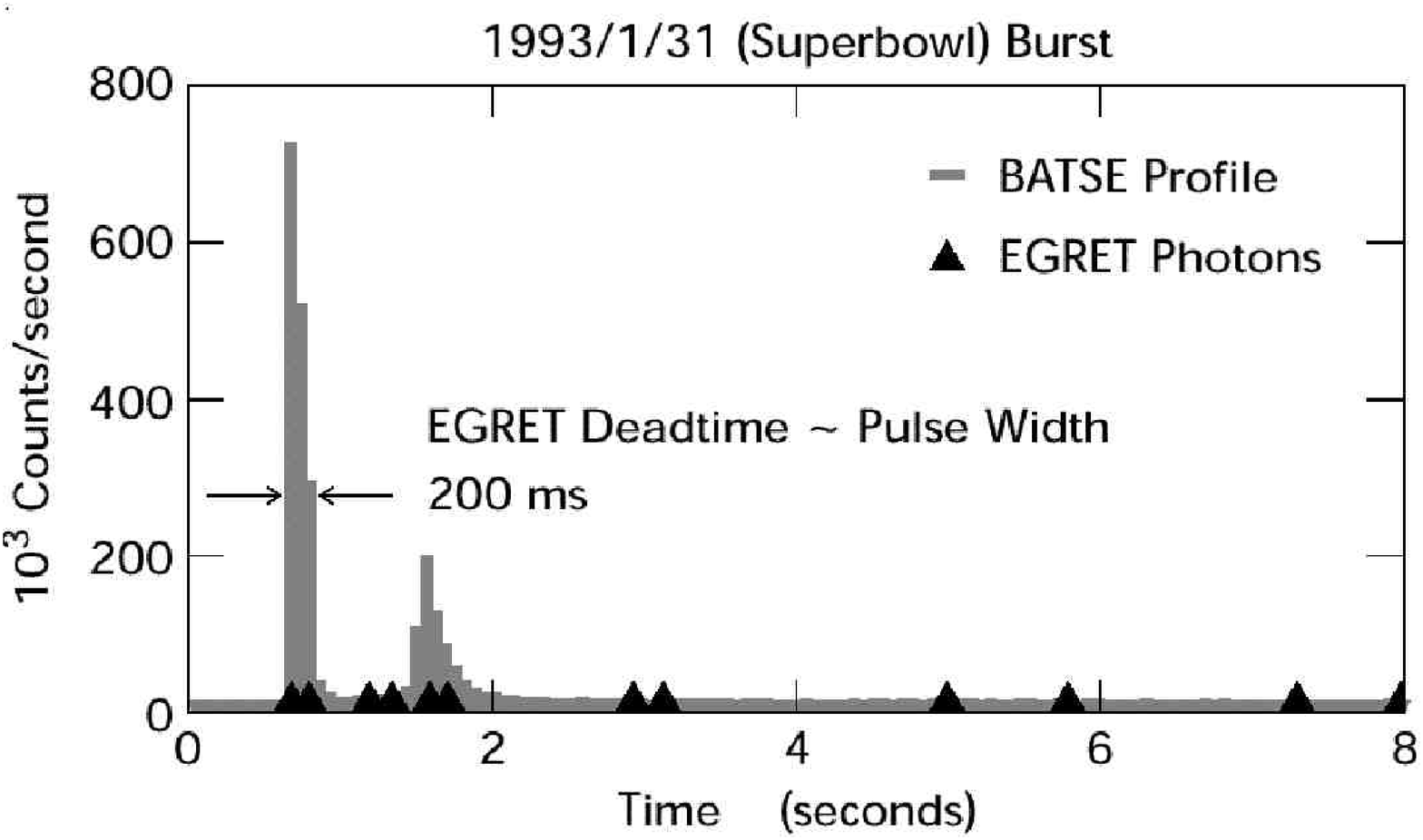,width=0.8\textwidth,angle=0,clip=}
\caption{\label{grbpuls} \it EGRET and BATSE light curves of the Superbowl burst, GRB930131.
The burst consisted of an extremely intense spike, followed by low-level
emission for several seconds. The true temporal development at energies $>$100 MeV is uncertain since EGRET dead time is comparable to
GRB pulse widths.}
\vspace{0.3cm}
%
    \epsfig{file=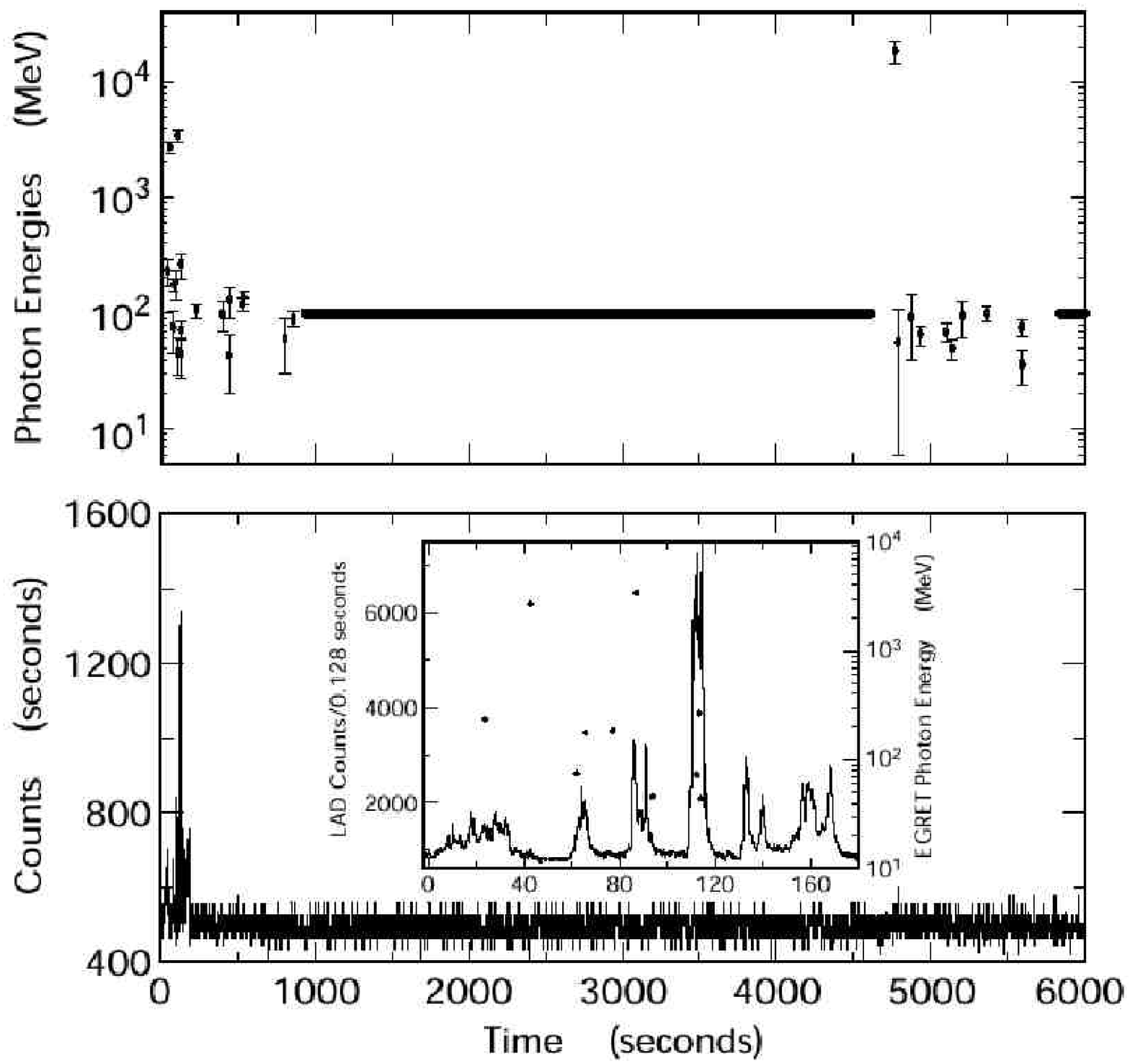,width=0.7\textwidth,angle=0,clip=}
\caption{\label{grb18} \it EGRET and BATSE light curves of GRB940217. Burst cessation at BATSE energies occurs
at ~160~s. Extended emission at EGRET energies persist beyond an
intervening earth occultation, up to 5000 seconds after the BATSE event. }
\end{figure}
%
\subsection{Gamma-ray burst}

Gamma-ray bursts (GRBs) are intermittently the most intense and most distant known sources of high-energy
gamma rays; at GeV energies, the brightest GRBs are 1000-10,000 times brighter than the brightest AGN. The
unparalleled luminosities and cosmic distances of GRBs, combined with their extremely fast temporal
variability, make GRBs an extremely powerful tool for probing fundamental physical processes and cosmic
history.

GLAST, in  concert with the Gamma-ray Burst Monitor, will measure the energy spectra of GRBs from a few keV
to hundreds of GeV during the short time after onset when the vast majority of the energy is released.
GLAST will also promptly alert other observers, thus allowing the observations of GLAST to be placed in the
context of multiwavelength afterglow observations, which are the focus of HETE-2 and the upcoming Swift
missions. The additional information available from GLAST's spectral variability observations will be key to
understanding the central engine.

 Figure~\ref{grbpuls} illustrates a very intense, short GRB. The true EGRET time profile is very uncertain
because the $\sim$ two hundred milliseconds EGRET dead time per photon is comparable to GRB pulse widths;
hence, many more photons may have been incident on EGRET during the extremely intense initial pulse.  The
GLAST dead time will be $\sim$ 10,000 times smaller, thus allowing  a precise measurement of the gamma-ray
flux during the peak. 

This characteristic together with its larger field of view and larger effective area,
should permit to detect virtually all
GRBs in its field of view reaching the "the edge" of the GRB distribution, as does BATSE.
Figure~\ref{grb18} shows another intense burst with very different temporal character
which occurred in EGRET's field of view on 1994 Feb 17.  At BATSE energies (25 - 1000 keV), this event
persisted for $\sim$160~s; however, at EGRET energies, it apparently continued at a relatively high flux
level past an Earth occultation, for at least 5000 s, to deliver a delayed $\sim~18$~GeV photon.
 GLAST, with negligible self veto, will have good efficiency above 10~GeV and it will be able to
localize GRBs with sufficiently high accuracy to enable rapid searches at all longer wavelengths. About half
of the 200 bursts per year detected by GLAST will be localized to better than 10~arc~minute radius, an
easily imaged field for large-aperture optical telescopes.
\begin{figure}[ht]
    \epsfig{file=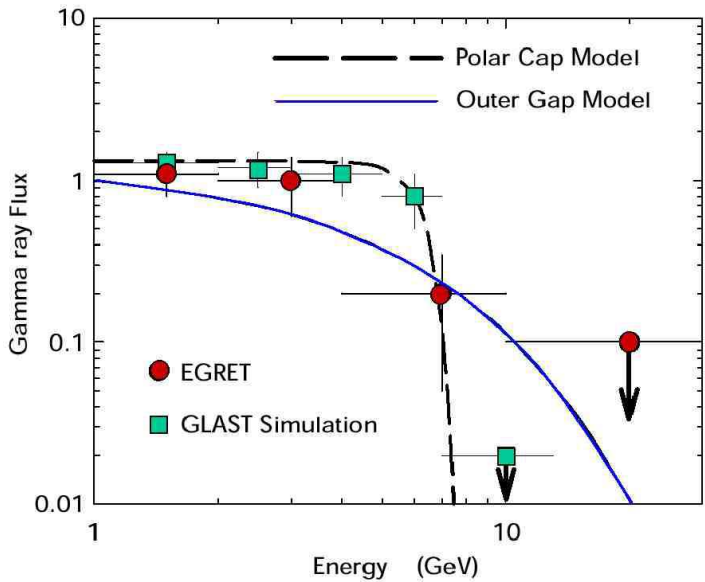,width=0.8\textwidth,angle=0,clip=}
\caption{\label{glastpuls} \it Modeled high-energy pulsar spectrum, showing the improvement in 
resolution between EGRET and GLAST. The polar cap model  predicts a sharp high-energy cutoff, 
while the outer gap model  predicts a more gradual cutoff. Unlike EGRET, GLAST will be able to
distinguish the true shape of the spectrum (assumed to be that of the polar
cap model in this simulation).}
\end{figure}
\subsection{Pulsars}
GLAST will discover many gamma-ray pulsars, potentially 50 or more, and will provide definitive
spectral measurements that will distinguish between the two primary models proposed to explain particle
acceleration and gamma-ray generation: the outer gap\cite{puls_outer} and polar cap models\cite{puls_polar}
 ( see figure~\ref{glastpuls}). 
From observations made
with  gamma ray experiments through the EGRET era, seven gamma-ray pulsars are known. GLAST will detect
more than 100 pulsars and will be able to directly search for periodicities in all EGRET unidentified sources.
Because the gamma-ray beams of pulsars are apparently broader than their radio beams, many radio-quiet,
Geminga-like pulsars likely remain to be discovered. 
\begin{figure}
\vspace{0cm}
    \epsfig{file=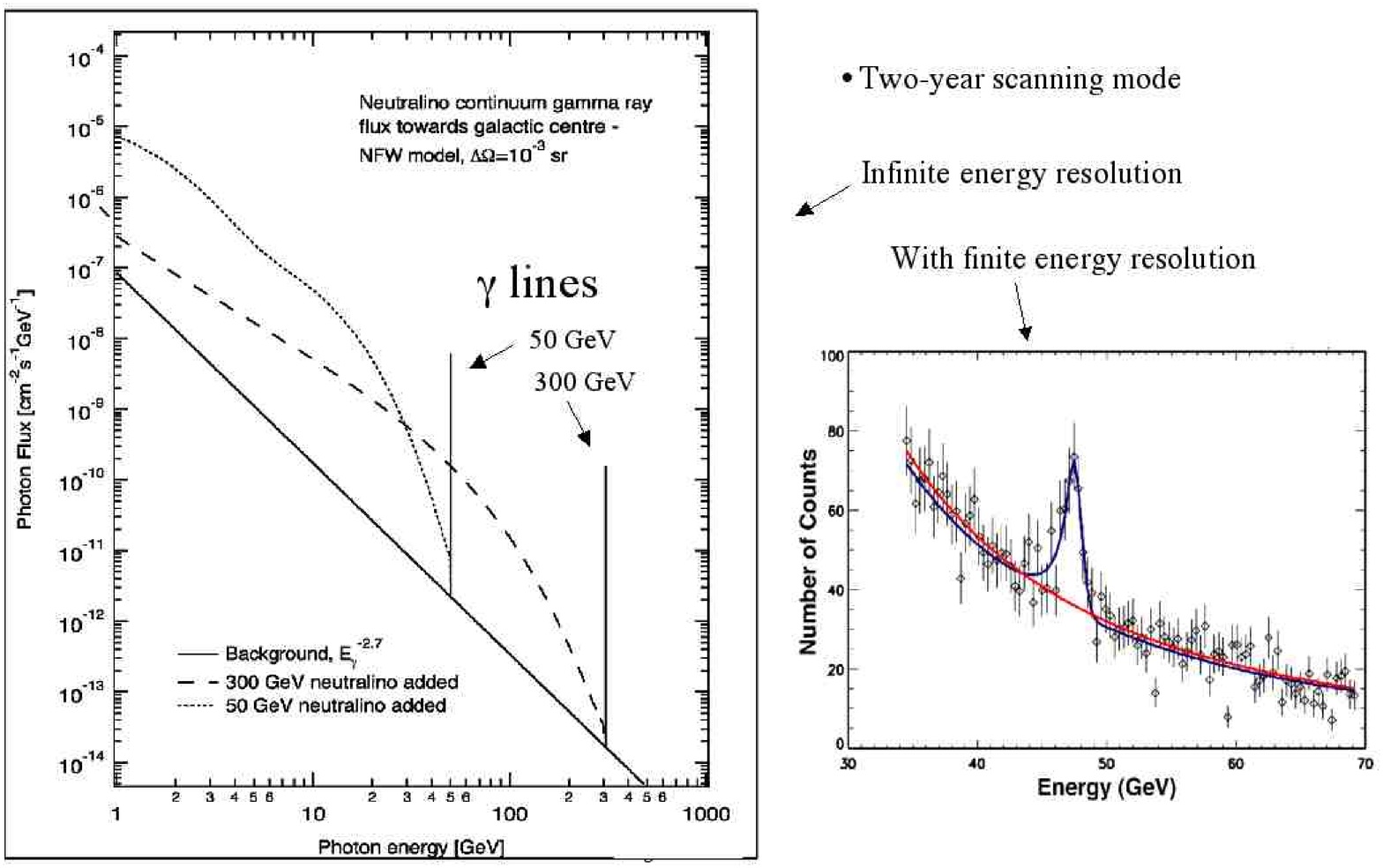,width=1\textwidth,angle=0,clip=}
\caption{\label{gline2} \it Total photon spectrum from the
galactic center from
$\chi \chi $ annihilation (on the left), and number of photons expected in GLAST for 
$\chi \chi  \rightarrow \gamma \gamma$ from a 1-sr cone near the galactic center with a 1.5 \% energy resolution (on the
right) }
\vspace{0.2cm}
    \epsfig{file=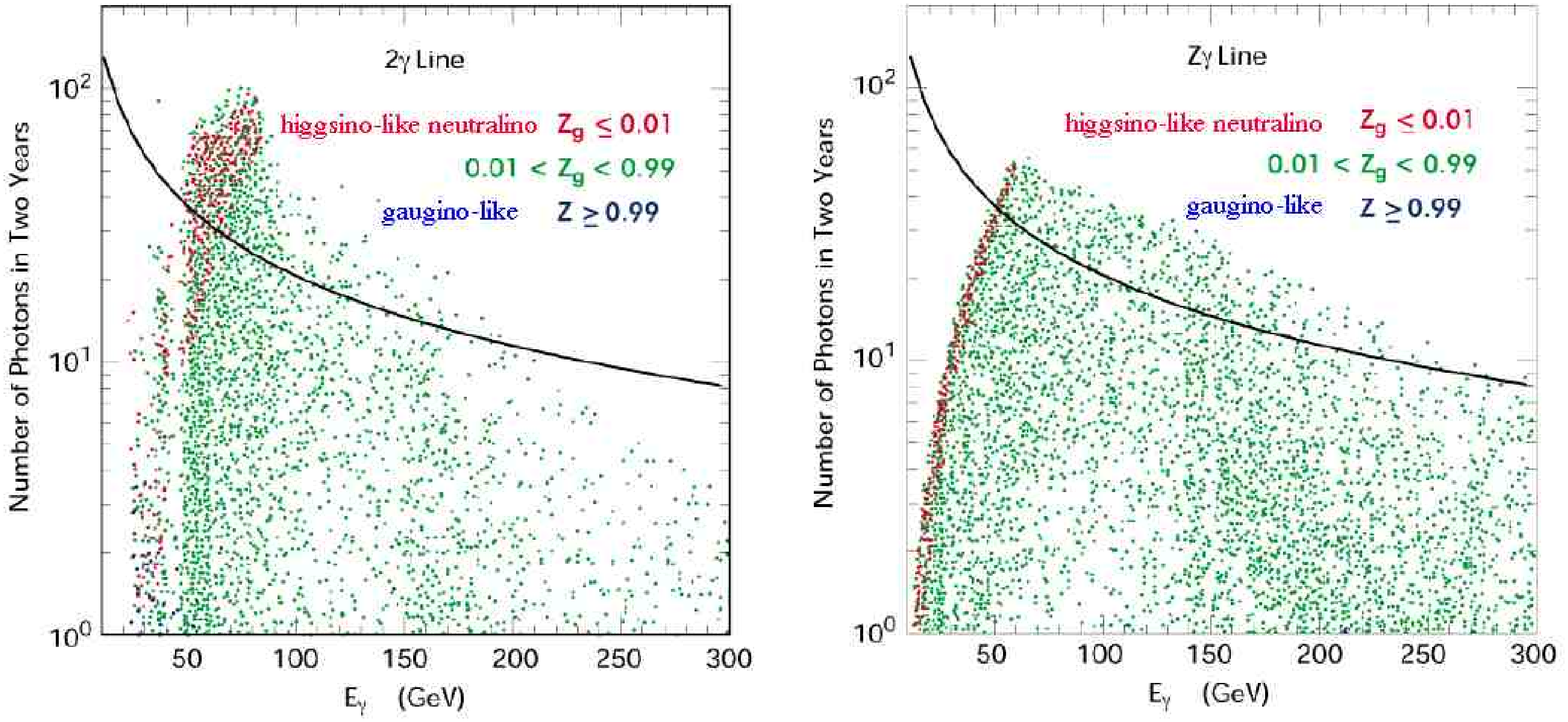,width=1\textwidth,angle=0,clip=}
\caption{\label{super_spa} \it Number of photons expected in GLAST for 
$\chi \chi  \rightarrow \gamma \gamma$ from a 1-sr cone near the galactic center as a function of the possible neutralino mass.
The solid line shows the number of events needed to obtain a five sigma signal detection over the galactic diffuse
gamma-ray background as estimated by EGRET data.\label{super_spa2}}
\end{figure}
\subsection{Search for supersymmetric dark matter}
\par\noindent GLAST is particularly interesting for the supersymmetric particle search  because,  
if neutralinos make up the dark matter of our galaxy, they would have non-relativistic
velocities, hence the neutralino annihilation into the gamma gamma  and gamma Z
final states can give rise to gamma rays with unique energies $E_\gamma= M_\chi$ and ${E_\gamma'=~M_\chi~(1-m_z^2/4M_\chi^2)}$.  

In figure~\ref{gline2} is shown how strong can be the signal\cite{Bergstro} in the case of a cuspy dark matter
halo profiles distribution\cite{navarro}.

Figure~\ref{super_spa} shows the GLAST capability to probe  the supersymmetric dark matter
hypothesis\cite{Bergstro}.  The various zone sample the MSSM
with different values of the parameters space  for  three classes of neutralinos.
 The previous galaxy dark matter halo profile\cite{navarro} that gives the maximal flux has been assumed. The
solid line shows the number of events needed to obtain a 5 $\sigma$ detection over the galactic diffuse
$\gamma$-ray background as estimated from EGRET data.  As the figures show,
a significant portion of the MSSM phase space is explored, particularly for the higgsino-like neutralino case. 

This effort will be complementary to a similar search for neutralinos looking with 
cosmic-ray experiments like the next space experiment PAMELA\cite{pam1} at the distortion of the secondary 
positron fraction and  secondary antiproton flux induced by a signal from a heavy neutralino. 
\begin{figure}
\vspace{0cm}
    \epsfig{file=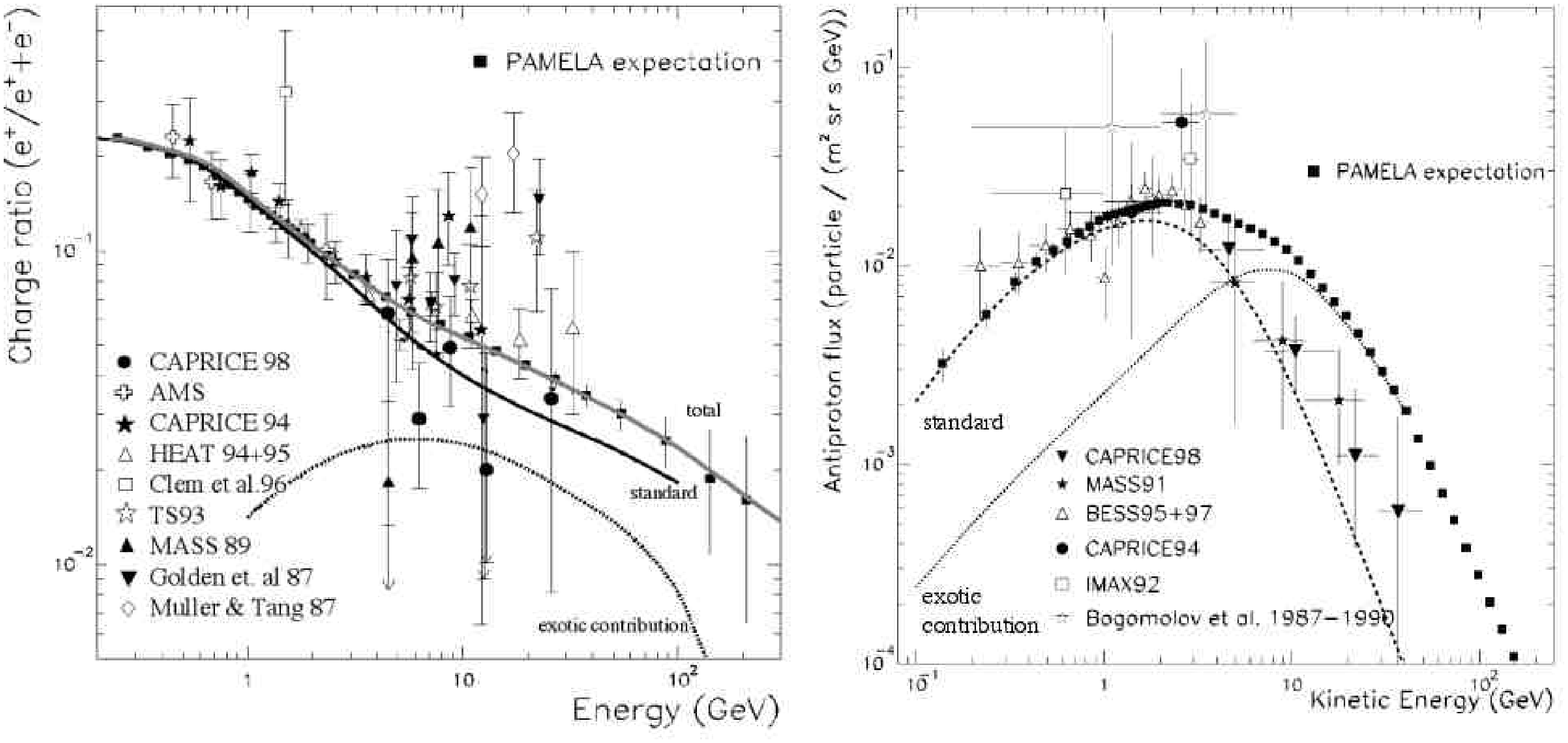,width=1\textwidth,angle=0,clip=}
\caption{\label{pamela2} \it  Distortion of the secondary 
positron fraction (on the left) and  secondary antiproton flux (on the right) induced by a signal from a
heavy neutralino. The PAMELA expectation in the case of  exotic contributions are shown by black squares} 
\end{figure}
\begin{figure}[ht]
\vspace{0cm}
    \epsfig{file=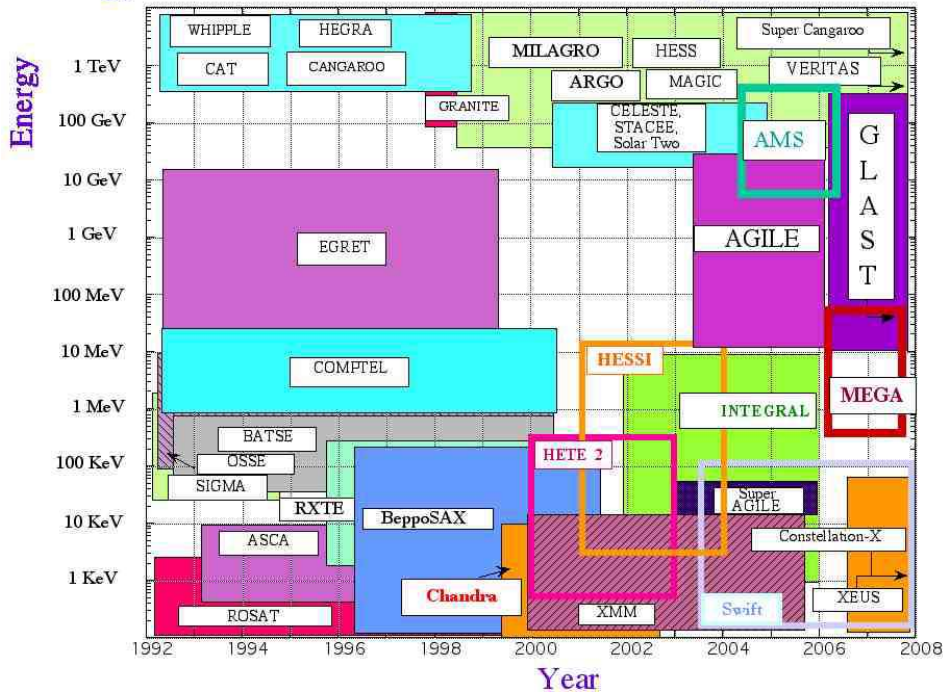,width=1.05\textwidth,angle=0,clip=}
\caption{\label{timeline2} \it  Timeline schedule versus the energy range covered by present and future
detectors in X and gamma-ray astrophysics.  }
\end{figure}

In figure~\ref{pamela2} (on the left) there are the experimental data\cite{el_data} for  the  positron
fraction  together with the distortion of the secondary positron fraction (solid line) due to
one possible contribution from neutralino annihilation (dotted line, from\cite{pam_e}).  
The expected data
from the experiment PAMELA  in the annihilation scenario for one year of operation are shown
by  black squares\cite{pamela}.

In the same figure~(on the right)
there are the  experimental data   for the antiproton flux\cite{antip_data} together with the 
distortion on the antiproton flux (dashed line) due to one possible
contribution from neutralino annihilation (dotted line, from\cite{Ullio}). 
 The antiproton data that PAMELA
would obtain in a single year of observation for
one of the Higgsino annihilation models are shown
by black squares.

\section{ Conclusion}

The gamma-ray space experiment  GLAST is under construction. Its time
of operation and  energy range is shown together with the other space X-ray satellite and gamma-ray
experiments in figure~\ref{timeline2}.  Note that it will cover an interval not covered by any other
experiments. Note also the number  of other experiments in other frequencies that will allow
extensive multifrequency studies.
\begin{figure}[ht]
\vspace{0cm}
    \epsfig{file=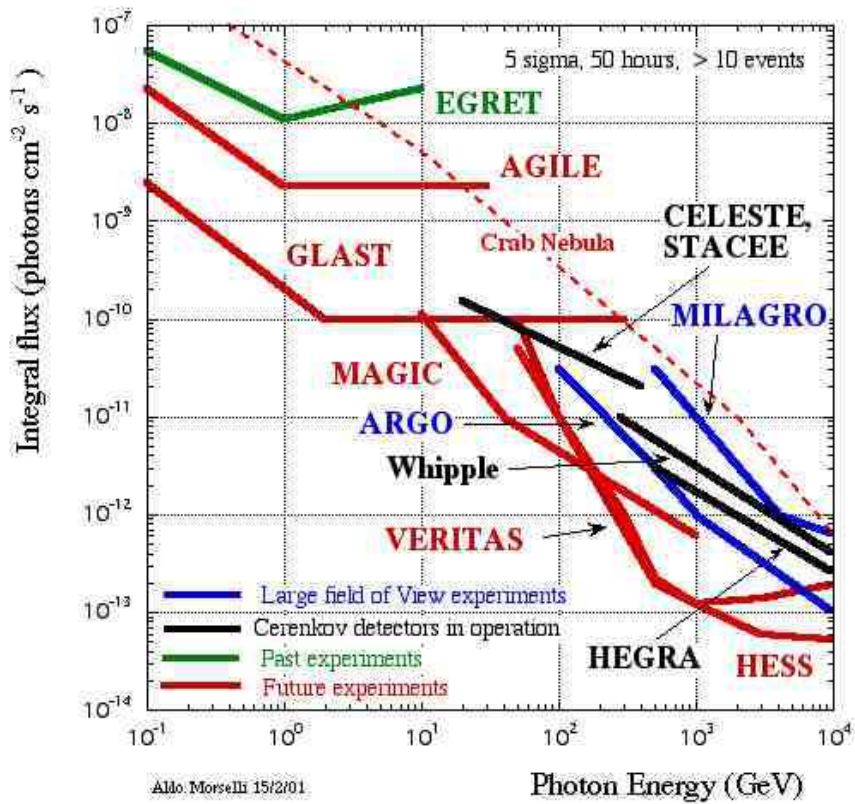,width=1\textwidth,angle=0,clip=}
\caption{\label{sens2} \it Sensitivity of present and future detectors in  the
gamma-ray astrophysics.  }
\end{figure}
%

In the last decade, ground-based instruments have made great progress, both
in technical and scientific terms. High-energy gamma rays can be observed from the ground by experiments
that detect the air showers produced in the upper atmosphere. Air shower
arrays directly detect the particles (electrons, muons, and photons) in air
showers, and atmospheric Cerenkov telescopes detect the Cherenkov radiation
created in the atmosphere and beamed to the ground. 
Detectors based on the atmospheric Cerenkov technique
consist of one or more mirrors that concentrate
the Cerenkov photons onto fast optical detectors. Photomultiplier
tubes (PMTs) placed in the focal plane are
generally used to detect the Cherenkov photons.
Two problems in using atmospheric Cherenkov telescopes
(ACT) are the night-sky background and the
large isotropic background from cosmic-ray showers.

The energy threshold of an atmospheric Cherenkov telescope
is determined by the number of Cherenkov photons
needed to observe a signal above the level of the
night-sky background.
For individual point sources, ground-based instruments have unparalleled
sensitivity at very high energies (above 50-250 GeV). 
For many objects, full multi-wave-length
coverage over as wide an energy range as possible will be needed to
understand the acceleration and gamma-ray production mechanisms.  On the
technical side, atmospheric Cherenkov telescopes have demonstrated that a
high degree of gamma/hadron discrimination and a source pointing accuracy
of 10-30 arc minutes (depending on the source strength) can be achieved
based on the detected Cherenkov image. Also the energy threshold is lowering 
remarkably (for a  review, see\cite{rev_gamma}).
 In figure~\ref{sens2}   the   GLAST sensitivity compared with the others present and future
detectors in the gamma-ray astrophysics  range is shown.
 The predicted sensitivity of a number of operational and proposed  Ground based Cherenkov telescopes,
 CELESTE, STACEE, VERITAS, Whipple is for a 50 hour exposure on a single source. EGRET, GLAST, MILAGRO, ARGO 
and AGILE sensitivity is shown for one year of all sky survey. The diffuse background assumed is
$2\cdot10^{-5} ~photons~cm^{-2} s^{-1} sr^{-1}(100 ~MeV/E)^{1.1}$, typical of the background seen by EGRET at high galactic latitudes.
The source differential photon number spectrum is assumed to have a power law index of -2, typical of many of the sources observed by EGRET and the
sensitivity is based on the requirement that the number of source photons detected is at least 5 sigma above the background.
  Note that on 
ground only MILAGRO and ARGO will observe more than one source simultaneously.  The Home Pages of the various instruments are at
{\sl http://www-hfm.mpi-hd.mpg.de/CosmicRay/CosmicRaySites.html}.
\begin{figure} [ht]
    \epsfig{file=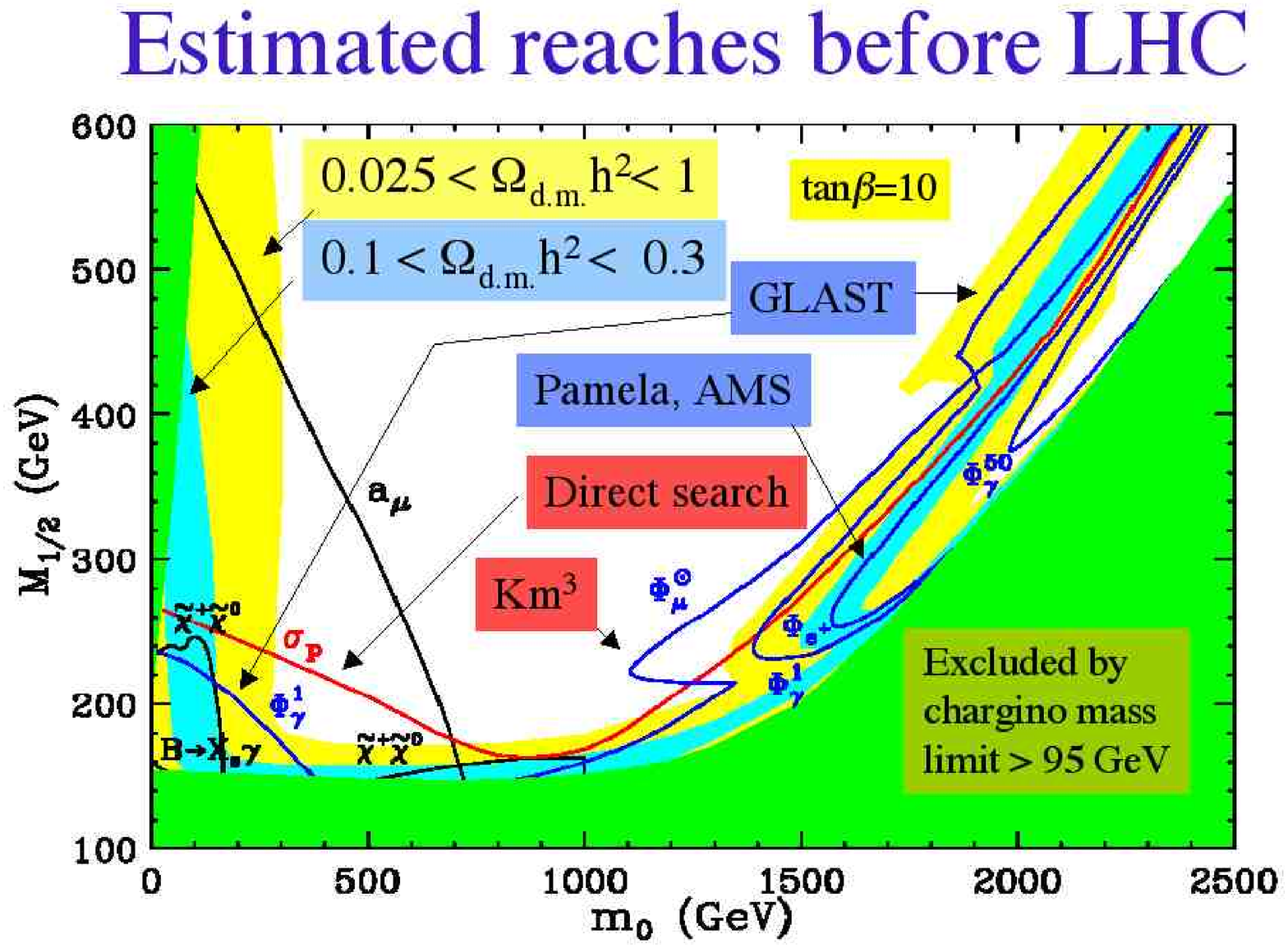,width=1\textwidth,angle=0,clip=}
\caption{\label{estim} \it Example of estimated reaches of various  searches  before the LHC begins
operation. Note the complementarity between the different techniques. For moderate values of $\tb$ 
all the cosmological interesting region will be covered (see text for details). } 
\end{figure}
As is shown in \cite{Feng}, a wide variety of experiments
provide interesting probes for the search of supersymmetric dark matter.
 In the next
five years, an  array of experiments will be sensitive to
the various potential neutralino annihilation products. These include
under-ice and underwater neutrino telescopes, atmospheric Cerenkov telescopes and the
 already  described space
 detectors  GLAST and  PAMELA together with  AMS. In many cases,
these experiments will improve current sensitivities by several orders of
magnitude and probably, as it is shown in \cite{Feng}, " all models with charginos or sleptons lighter than 300 GeV will produce observable signals in at least one
experiment in the cosmologically preferred regions of parameter space with $0.1 <
\Omegachi h^2 < 0.3$ " before LHC.
\begin{figure}
\vspace{0cm}
    \epsfig{file=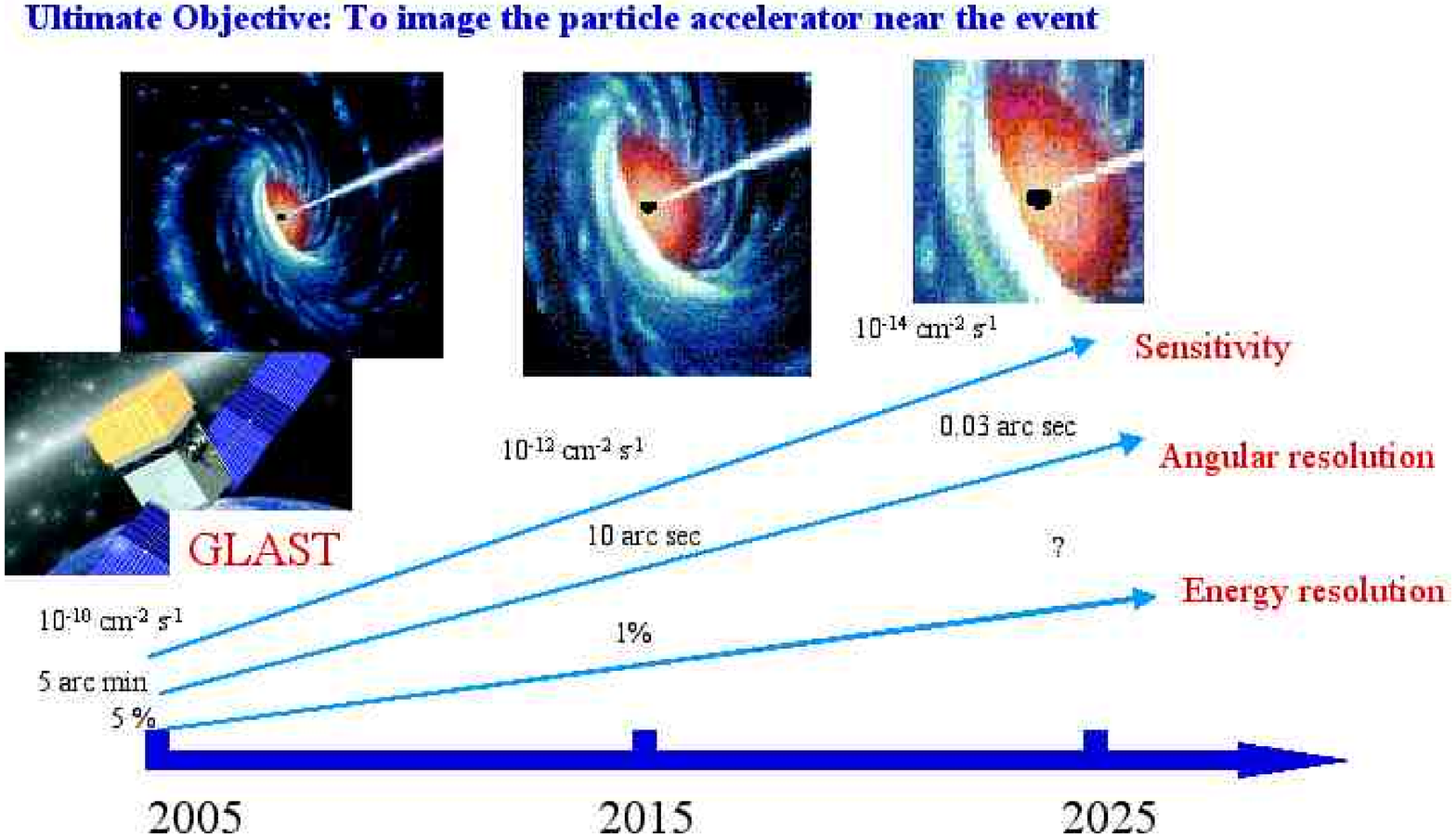,width=0.9\textwidth,angle=0,clip=}
\caption{\label{ultim2} \it Gamma-Ray Astronomy Long Term Plan }
\end{figure}
An example is shown in figure~\ref{estim}  in  the framework
of minimal supergravity, which is fully specified by the five
parameters (four continuous, one binary)
$m_0, \mgaugino, A_0, \tb, \sign (\mu) $.
Here, $m_0$, $\mgaugino$, and $A_0$ are the universal scalar mass,
gaugino mass, and trilinear scalar coupling. 
The figure shows the limits that can be obtained in the $m_0, \mgaugino $ plane
 for $\tb=10, ~A_0=0, ~\mu > 0 $. 
Higher values ($\sim$ 50 ) of $\tb$ requires significant fine-tuning of the electroweak
scale. The limit from gamma-ray assumes a moderate halo profile.

The a$_\mu$ curve refers to the expected region that will be probed before 2006 by
the measurements of the muon magnetic dipole moment\cite{a_mu}. The curve $B \to
X_s \gamma$ refers to the improvement expected for the same date from BaBar, BELLE
and B factories in respect to the CLEO and ALEPH results\cite{Aleph}. 
The curve $\Phi_{\mu}^{\odot}$ refers to the indirect DM search with underwater $\nu$ experiments like
AMANDA, NESTOR and ANTARES\cite{AMANDA_like} and the curve $\sigma_p$ refers to the direct DM search
with underground experiments like DAMA, CDMS, CRESST and GENIUS\cite{DAMA_like}

We conclude with one last remark, the angular resolution and energy resolution achievable in gamma ray
astrophysics  is still lower to what is desirable and achievable in other band; so a long term plan like the
one sketch in figure~\ref{ultim2} is needed and can bring spectacular results.


\section{Acknowledgments}
I wish to thanks all the participants to the school. 
Everybody (both professors and students)
 contributes  so much to discussions and debates to make the time spent in L'Aquila 
really very interesting.


\end{document}